\documentclass[showpacs, pra,onecolumn,preprintnumbers ,amsmath, amssymb, superscriptaddress, aps]{revtex4-2}
\usepackage{color}
\usepackage{amsmath,amssymb}
\usepackage{pifont}
\usepackage{amssymb}  
\usepackage{bbold}
\usepackage{float}
\usepackage{subfloat}

\usepackage[caption=false]{subfig}
\usepackage{tikz}
\usepackage{makecell}
\usepackage{subfig}
\usepackage{pifont}   
\usepackage{graphicx} 
\graphicspath{{Figures/}}
\usepackage{dcolumn}  
\usepackage{bm}       
\usepackage{multirow} 
\usepackage{placeins}
\usepackage[colorlinks]{hyperref}
\usepackage{mathtools}
\usepackage{appendix}

\begin{document}
	
	\title{Group delay time of fermions in graphene through tilted potential barrier}
	\date{\today}
	\author{Youssef Fattasse}
	\affiliation{Laboratory of Theoretical Physics, Faculty of Sciences, Choua\"ib Doukkali University, PO Box 20, 24000 El Jadida, Morocco}
	
	\author{Miloud Mekkaoui}
	\affiliation{Laboratory of Theoretical Physics, Faculty of Sciences, Choua\"ib Doukkali University, PO Box 20, 24000 El Jadida, Morocco}
	\author{Ahmed Jellal}
	\email{a.jellal@ucd.ac.ma}
	\affiliation{Laboratory of Theoretical Physics, Faculty of Sciences, Choua\"ib Doukkali University, PO Box 20, 24000 El Jadida, Morocco}
	\affiliation{Canadian Quantum  Research Center,
		204-3002 32 Ave Vernon,  BC V1T 2L7,  Canada}
	\author{Abdelhadi Bahaoui}
	\affiliation{Laboratory of Theoretical Physics, Faculty of Sciences, Choua\"ib Doukkali University, PO Box 20, 24000 El Jadida, Morocco}
	
	
	\begin{abstract}
		
		The group delay time of Dirac fermions subjected to  a tilting barrier potential along the $ x $-axis  is investigated in graphene. 
		We start by finding the eigenspinor solution of the Dirac equation and then relating it to incident, reflected, and transmitted beam waves. 
	 This relationship allows us to compute the group delay time in transmission and reflection by obtaining the corresponding phase shifts. 
		We discovered that the barrier width, incident energy, and incident angle can all be used to modify the group delay time, and that the particles travel through the barrier at the Fermi velocity $ v_F $. 
		Our findings also show that the transmission group delay might be controlled, and that gate voltage control could be useful in graphene-based tilting barriers. 

	\end{abstract}

	\maketitle

\section{Introduction}
The ability to change particle behavior by adjusting the gate bias voltage has prompted a lot of interest in quantum tunneling
\cite{Novoselov, Zhang,Nilsson}. 
Electrical conduction in gapless graphene cannot be turned off using the control voltages required for conventional transistor operation  \cite{Katsnelson}. 
This problem can be solved in graphene by establishing an energy gap in the energy spectrum of particles. 
The desired gap is a measurement of the threshold
voltage and the on-off ratio of the field effect transistors \cite{Lin,Kedzierski}. 
However, significant progress has been made in understanding quantum phenomena in graphene systems, with the group delay time being one of the most important parameters connected to the dynamic aspect of the tunneling process \cite{Hartman, Zhenhua}. By studying the behavior of wave packets,
Hartman demonstrated that the group delay may be described in terms of the derivative of the phase shift with respect to the energy \cite{Hartman}. 
The Hartman effect states that the effective group velocity of a particle can become superluminal for sufficiently large barriers  \cite{Olkhovsky, Zhenhua}. 
%
%
%


We previously investigated the  quantum tunneling for Dirac fermions in graphene scattering by a linear vector potential 
 \cite{HBahlouli, Mekkaoui19}. In   \cite{HBahlouli}, it was demonstrated that 
the infinite mass boundary condition  discretizes  the transverse momentum. 
As a result, an effective massive 1D Dirac equation is derived in which the quantized transverse momentum behaves as an effective mass.
 In \cite{Mekkaoui19},
the Goos-H\"anshen shifts were studied using the solutions of the energy spectrum of  graphene in a linear barrier potential.
The procedure begins with the transmission and reflection probabilities being used to determine the appropriate phase shifts.
According to a numerical analysis, incident energy, barrier height, and width have a significant impact on the Goos-Hänshen shifts, which can change positively or negatively under particular conditions.    

We investigate the group delay time in transmission and reflection for Dirac fermions in graphene subjected  to a tilting barrier potential based on our prior work \cite{HBahlouli, Mekkaoui19}. We illustrate how to calculate the group delay time as a function of different  physical parameters based on phase shifts and GH shifts in transmission using the energy spectrum solution. We propose a numerical investigation under various conditions  to provide a better understanding of our findings. In particular,
the tilting barrier is shown to be able to manage the group delay time.

The following is a breakdown of the current paper's structure. 
We define our problem in section \ref{ensp}, write out the corresponding Hamiltonian and energy spectrum solutions for different regions. We then calculate the transmission and reflection probabilities from which the phase shifts are determined.  
As a result, we determine the group delay time in terms of the physical parameters that define our system using traditional definitions in section \ref{gdtime}. 
We numerically examine and highlight the basic aspects of the group delay time in section \ref{numan}. In the concluding section, we summarize our findings.

\section{Energy spectrum}\label{ensp}

\begin{figure}[H]
	\centering{
		\includegraphics[scale=0.6]{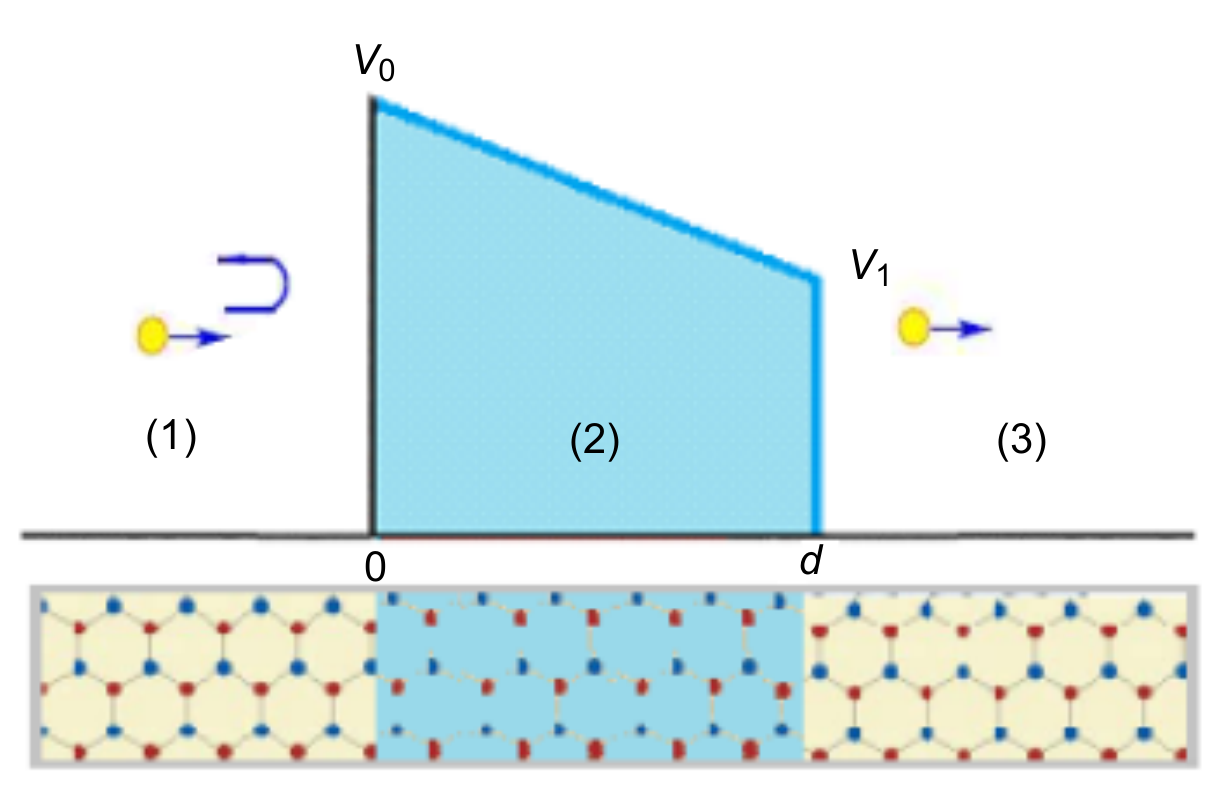}
	}
	\caption{(color online) {Configuration of a tilted barrier potential
			of width $d$, height $V_0$ and $V_1$  applied to an intermediate region of  graphene.}}
	\label{db.1}
\end{figure}

As seen schematically in Fig. \ref{db.1}, massless Dirac fermions in graphene are scattered by tilted  barrier potentials $V_0$ and $V_1$.
The current system is divided into three zones designated by the numbers $ j = 1, 2, $ and 3, each of which has a distinct potential.  

The following Dirac-like Hamiltonian can be used to describe the current system
\begin{equation}\label{Ham1}
H=v_{F} {\boldsymbol{\sigma}}\cdot\textbf{p}-\left(\beta
x-V_0\right)\Theta(xd-x^{2}){\mathbb I}_{2}
\end{equation}
with  the Heaviside step function $\Theta$, the Fermi velocity  ${v_{F}\approx 10^6
m/s}$, the Pauli
matrices
${{\boldsymbol{\sigma}}=(\sigma_{x},\sigma_{y})}$, $\textbf{p}=-i\hbar(\partial_{x},
\partial_{y})$,  the $2 \times 2$ unit matrix ${\mathbb I}_{2}$, $\beta=\frac{V_0-V_1}{d}$.
The spinor $\Psi(x,y) $ at energy $E$ has a time-independent Dirac equation, which is given by 
\begin{equation} \label{eqh1}
\left[v_{F} {\boldsymbol{\sigma}}\cdot\textbf{p}-\left(\beta
x-V_0\right)\Theta(xd-x^{2}){\mathbb I}_{2}\right]\Phi(x,y)=E
\Phi(x,y).
\end{equation}
The system is considered to have a finite width $ W $.
The spinor meets the infinite mass boundary condition  at the interfaces $ y = 0 $ and $ y = W $ along the $ y $-direction
\cite{Tworzydlo, Berry}.
As a result, the transverse momentum $k_{y}$ is quantized
\begin{equation}
k_{y}=\frac{\pi}{W}\left(n+\frac{1}{2}\right),\qquad n=0,1,2 \cdots.
\end{equation}

The spinor is then obtained by separating  the variables and, therefore, we write
$\Psi_{j}(x,y)=\left(\varphi_{j}^{+}(x),\varphi_{j}^{-}(x)\right)^{\dagger}e^{ik_{y}y}$. 
In region 1 ($x < 0$), we determine the two components of the eigenspinor after solving the eigenvalue equation 
\begin{eqnarray}\label{eq3}
   \Psi_{1} (x,y)=  \left(
            \begin{array}{c}
              {1} \\
              {z_{1}} \\
            \end{array}
          \right) e^{i(k_{1}x+k_{y}y)} + r\left(
            \begin{array}{c}
              {1} \\
               {-z_{1}^{-1}} \\
            \end{array}
          \right) e^{i(-k_{1}x+k_{y}y)}, \qquad z_1 
          =s_{1}e^{i\phi}
\end{eqnarray}
 where $r$ is the reflection coefficient and $\phi=\tan^{-1}(k_{y}/k_1)$ is
 the incident angle, while the sign function $s_{1}={\mbox{sign}}{\left(E\right)}$ 
 indicates the conduction and valence bands. The corresponding dispersion relation is straightforward to obtain
 \begin{align}
 E=s_1\hbar v_{F}\sqrt{ k_1^2 +k_y^2}. 
 \end{align}

In region 2 ($0<x<d$), the parabolic cylinder function  can be used to express the general solution, 
with the two components being  \cite{HBahlouli}
\begin{align}\label{hii1}
& \Xi^{+}=b_{1}
 D_{\eta-1}\left(\Lambda\right)+b_{2}
 D_{-\eta}\left(-\Lambda^{*}\right)
\\
&\label{hii2}
\Xi^{-}=-\frac{b_{1}}{k_{y}}\sqrt{2\varrho}e^{-i\pi/4}
 D_{\eta-1}\left(\Lambda\right)-\frac{b_{2}}{k_{y}}\left[ 2(\epsilon- \varrho x)
 D_{-\eta}\left(-\Lambda^{*}\right)
+
 \sqrt{2\varrho}e^{i\pi/4}D_{-\eta+1}\left(-\Lambda^{*}\right)\right]
\end{align}
and we have defined
$\eta=\frac{ik_{y}^{2}}{2\varrho}$,
$\Lambda(x)=\sqrt{\frac{2}{\varrho}}e^{i\pi/4}\left(-\varrho
x+\epsilon\right)$, $\varrho=\frac{\beta}{\hbar v_{F}}$, 
$\epsilon=\frac{E-V_{0}}{\hbar v_{F}}$, with $b_{1}$ and $b_{2}$ are two constants. 
The following are the components of the spinor solution of the Dirac equation \eqref{eqh1}  in region 2 
\begin{align}
\varphi^{+}(x)=\Xi^{+}+i\Xi^{-}, \qquad
\varphi^{-}(x)=\Xi^{+}-i\Xi^{-}
\end{align}
which results in the spinor 
\begin{eqnarray}
 \Psi_{2 } (x,y) &=& a_{1}\left(%
\begin{array}{c}
 \chi^{+}(x) \\
  \chi^{-}(x) \\
\end{array}%
\right)e^{ik_{y}y}+a_{2}\left(%
\begin{array}{c}
 \xi^{+}(x) \\
 \xi^{-}(x)\\
\end{array}%
\right)e^{ik_{y}y}
\end{eqnarray}
 where the functions $ \chi^{\pm}(x)$ and $\xi^{\pm}(x)$
 are written as follows: 
\begin{eqnarray}
\chi^{\pm}(x)&=&
 D_{\eta-1}\left(\Lambda\right)\mp
 \frac{1}{k_{y}}\sqrt{2\varrho}e^{i\pi/4}D_{\eta}\left(\Lambda\right)\\
\xi^{\pm}(x)&=&
 \pm\frac{1}{k_{y}}\sqrt{2\varrho}e^{-i\pi/4}D_{-\eta+1}\left(-\Lambda^{*}\right)
  \pm
 \frac{1}{k_{y}}\left(-2i\epsilon_{0}\pm
 k_{y}+2i \varrho x\right)D_{-\eta}\left(-\Lambda^{*}\right)
\end{eqnarray}
$a_1$ and $a_2$ are two constants.
 
The following spinor is found in region 3 $ (x > d) $: 
\begin{equation}\label{eq6}
 \Psi_{3} (x,y)= t \left(
            \begin{array}{c}
              {1} \\
              {z_{1}} \\
            \end{array}
          \right) e^{i(k_{1}x+k_{y}y)}
\end{equation}
propagating with the same wave vector $ k_1 $ as in region 1.
 The  transmission $ t $ and reflection $ r $ coefficients associated with phase shifts will be computed using the previous solutions. 

\section{Group delay time}\label{gdtime}

Let us begin by defining the following abbreviations in order to determine the group delay time. 
This is about the eigenspinors' components
\begin{equation}
\chi^{\pm}(0)=\chi_{0}^{\pm},\qquad
 \chi^{\pm}(d)=\chi_{d}^{\pm},\qquad
 \xi^{\pm}(0)=\xi_{0}^{\pm},\qquad
 \xi^{\pm}(d)=\xi_{d}^{\pm}.
\end{equation}
Because the spinors must be consistent at all interfaces,  
we get a set of equations stated in terms of transfer matrices $M_{jj+1}$ between different regions. Then, over the entire tilted barrier, the full transfer matrix can be expressed as 
\begin{equation}
\label{systm1}
\left(%
\begin{array}{c}
  1 \\
  r \\
\end{array}%
\right)=M\left(%
\begin{array}{c}
  t \\
  0 \\
\end{array}%
\right)=\prod_{j=1}^{4}M_{j j+1}\left(%
\begin{array}{c}
  t \\
  0 \\
\end{array}%
\right)
\end{equation}
where $M_{12}$, $M_{2 3}$ are transfer matrices that connect the $j$-th region wavefunction to the $(j +
1)$-th region wavefunction. Explicitly, we have
\begin{align}
M_{12}=\left(%
\begin{array}{cc}
  1 &1 \\
  z_{1} & -z^{\ast}_{1}  \\
\end{array}%
\right)^{-1}\left(%
\begin{array}{cc}
\chi_{0}^{+} &  \xi_{0}^{+}\\
 \chi_{0}^{-} & \xi_{0}^{-}\\
\end{array}%
\right),
\quad
M_{23}=\left(%
\begin{array}{cc}
 \chi_{d}^{+} &  \xi_{d}^{+}\\
 \chi_{d}^{-} & \xi_{d}^{-} \\
\end{array}%
\right)^{-1}\left(%
\begin{array}{cc}
  e^{\textbf{\emph{i}}k_{1} d} & e^{-\textbf{\emph{i}}k_{1} d} \\
  z_{1} e^{\textbf{\emph{i}}k_{1} d}  & -z_{1}^{\ast} e^{-\textbf{\emph{i}}k_{1} d}  \\
\end{array}%
\right)
\end{align}
and then
\begin{align}
	M=\left(%
	\begin{array}{cc}
		m_{11} & m_{12} \\
		m_{21} & m_{22} \\
	\end{array}%
	\right).
\end{align}
As a result, the transmission and reflection coefficients are given by 
\begin{align}
t=\frac{1}{|m_{11}|} e^{i\varphi_{t}}, \qquad
r=\left|\frac{m_{21}}{m_{11}}\right|e^{i\varphi_{r}}
\end{align}
where
\begin{align}
\varphi_{t}=\arctan\left(i\frac{t^{\ast}-t}{t+t^{\ast}}\right), \qquad
\varphi_{r}=\arctan\left(i\frac{r^{\ast}-r}{r+r^{\ast}}\right)	
\end{align}
are the phase shifts of the transmission and reflection
amplitudes, respectively. After some lengthy but straightforward algebra, we obtain the following coefficients:
\begin{eqnarray}
 t=\frac{e^{-ik_{1}d}\left(1+z_{1}^{2}\right)\left(\xi^{+}_{d}\chi^{-}_{d}-\xi^{-}_{d}\chi^{+}_{d}\right)}
{\vartheta},\qquad r=\frac{\delta} {\vartheta}
\end{eqnarray}
as well as having defined
\begin{eqnarray}
\delta&=&\chi^{+}_{0}\xi^{-}_{d}+z_1\left(\chi^{-}_{d}\xi^{-}_{0}+\chi^{+}_{0}\xi^{+}_{d}-\chi^{+}_{d}\xi^{-}_{0}
+\chi^{+}_{0}\xi^{-}_{d}\right)-z_{1}^{2}\left(\chi^{+}_{d}\xi^{+}_{0}-
\chi^{+}_{0}\xi^{+}_{d}-\chi^{-}_{d}\xi^{+}_{0}\right)\\
\vartheta&=&\left(\xi^{+}_{0}+z_1\xi^{-}_{0}\right)\left(\chi^{-}_{d}-z_1\chi^{+}_{d}\right)-
\left(\chi^{+}_{0}+z_1\chi^{-}_{0}\right)\left(\xi^{-}_{d}-z_1\xi^{+}_{d}\right).
\end{eqnarray}
The transmission $T=\frac{J_t}{J_i}$ and reflection $R=\frac{J_r}{J_i}$ probabilities are calculated using the current of densities  $J_i$, $J_r$, and $J_t$ representing the incident, reflected and transmitted waves, respectively. 
 We get the current density from the Hamiltonian 
\begin{align}
	J=e\upsilon_F \psi^+\sigma_x\psi
\end{align}
leading to the probabilities
\begin{align}
T=\left|t\right|^{2}, \qquad 	R=\left|r\right|^{2}.
\end{align}	

A transverse wave vector $ ky = k_{y_0} $ and an incident angle
$\phi (k_{y_0})\in [0, \frac{\pi}{2}],$ indicated by the subscript 0, are then used to analyze the GH shifts and group delay time. 
A temporal-spatial wave packet, which is the weighed superposition of plane wave spinors, can be used to describe a finite pulsed electron beam.
As a result, the incident, reflected at $ x = 0 $, and transmitted wave packets at $ x = d $ wave functions can be expressed as a double Fourier integral over $ k_y $ \cite{Chen3x, Chen4x}
 \begin{align}
 &	\label{int1}
 	\psi_{i}(x,t)=\iint f(k_y,\omega) e^{-i(k_y y-\omega
 		t)}dk_yd\omega\\
 	&\label{int2}
 	\Psi_{r}(x,t)=\iint r f(k_y,\omega) e^{-i(k_y y-\omega
 		t)}dk_yd\omega
 	\\
 	& \label{int3}
 	\Psi_{t}(x,t)=\iint t f(k_y,\omega) e^{-i(k_y y-\omega
 		t)}dk_yd\omega
 \end{align}
where the involved spinors are solutions of Dirac equation \eqref{eqh1}. Here, $f(k_y,\omega)=w_ye^{-w_{y}^2(k_y-\omega)^2}$ is
the Gaussian angular spectral distribution, with $\omega=E/\hbar$ and $w_y$ is the
half beam width at waist \cite{Beenakker}. 
The total phases of reflected and transmitted waves at   $ x = 0, d $, respectively, are given by
\begin{align}
	\Phi_{\nu}=\varphi_{\nu}+k_y-\omega t, \qquad \nu=t, r. 	
\end{align}
Using the stationary phase approximation, we may get analytical equations for group delay and lateral GH shift by assuming that the distribution $f(k_y,\omega)$ is a smooth and steeply peaked function around the center energy/wavevector \cite{Steinberg1, Li1}. 
GH shifts are obtained from
 \begin{equation}
        S_{\nu}=- \frac{\partial \varphi_{\nu}}{\partial.
        k_{y}}.
 \end{equation}
For retaining the nice shape throughout propagation, the equation of motion is determined using the constraint $\partial\Phi_\nu/\partial\omega=0$. 
This provides the group delay time
\begin{align}
       \tau_{\nu}&= \frac{\partial \varphi_{\nu}}{\partial
        \omega}+\frac{\partial k_y}{\partial
        \omega}S_{\nu}\\
    &=\tau_{\nu}^{s}+\tau_{\nu}^{\varphi}   
 \end{align}
where
$\tau_{\nu}^{\varphi}$ represents  the time derivative of phase
shifts 
\begin{equation}
	\tau_{t}^{\varphi}=\hbar \frac{\partial \varphi_t}{\partial
		E}+\frac{\hbar}{2}\frac{\partial \phi}{\partial E}, \qquad
	\tau_{r}^{\varphi}=\hbar \frac{\partial \varphi_r}{\partial E}
\end{equation}
and the second $\tau_{\nu}^{s}$ results from the contribution
of  $S_{\nu}$
\begin{equation}
	\tau_{t}^{s}=\frac{\sin\phi}{\upsilon_F}S_t, \qquad
	\tau_{r}^{s}=\frac{\sin\phi}{\upsilon_F}S_r.
\end{equation}

We shall proceed with numerical analysis after obtaining closed form equations of the group delay in various energy domains.

\section{Numerical analysis}\label{numan}

We compute the group delay time in transmission for electrons passing through a tilting barrier under various incident angle $\phi$, potential height $V_0$ and width $d$, and incident energy $E$.
Dimensionless group delay time $\tau_{t}/\tau_{0}$, which results in transversal time $\tau_{0}=\frac{d}{v_F}\cos\phi$, is convenient for our task. The key findings of this study are depicted in the following seven figures, each with a distinct set of physical parameters.

\begin{figure}[H]
    \centering
    \subfloat[]{
        \centering
        \includegraphics[scale=0.48]
        {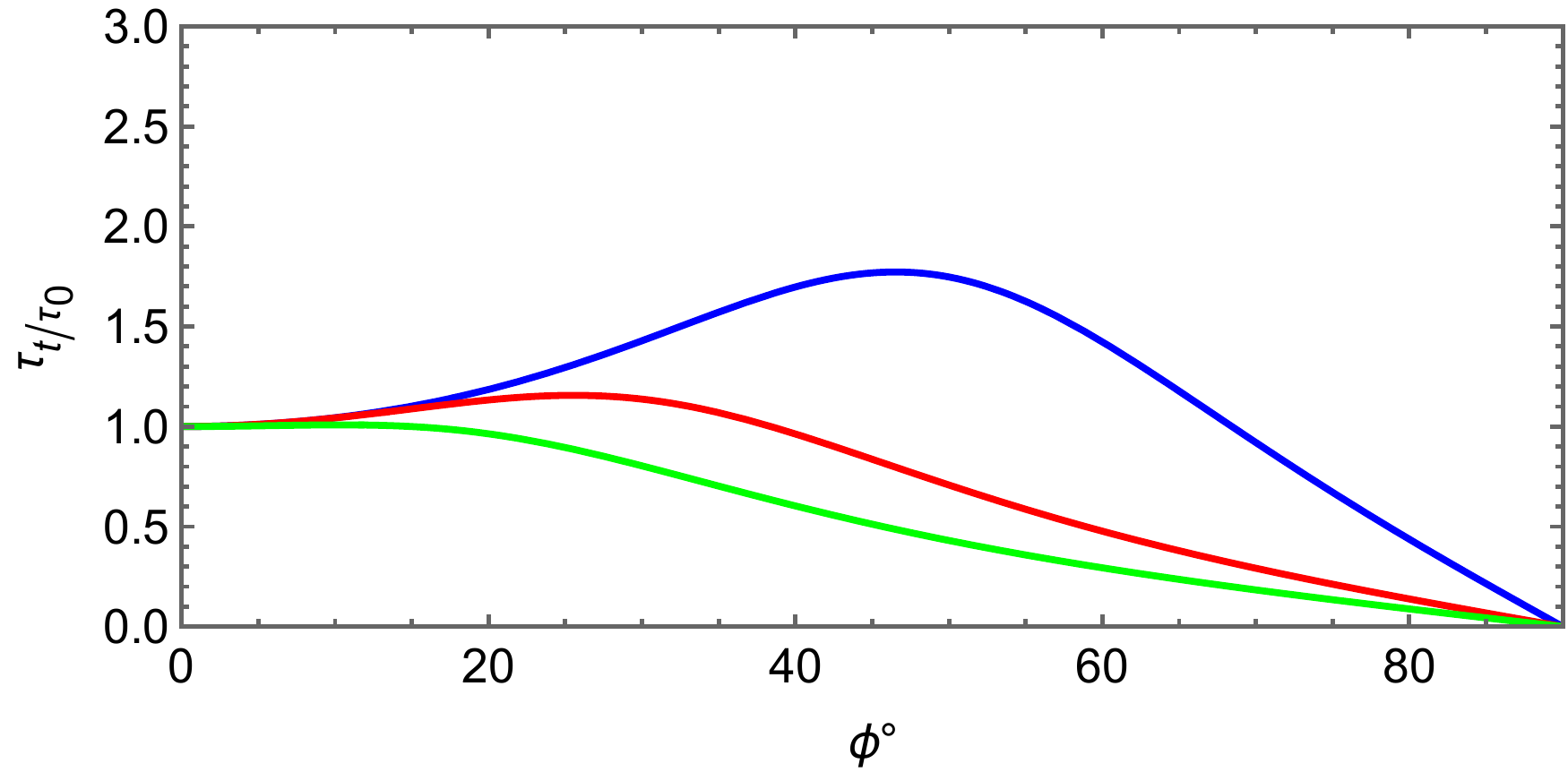}}
  \subfloat[]{
  	\centering\includegraphics[scale=0.48]
  	{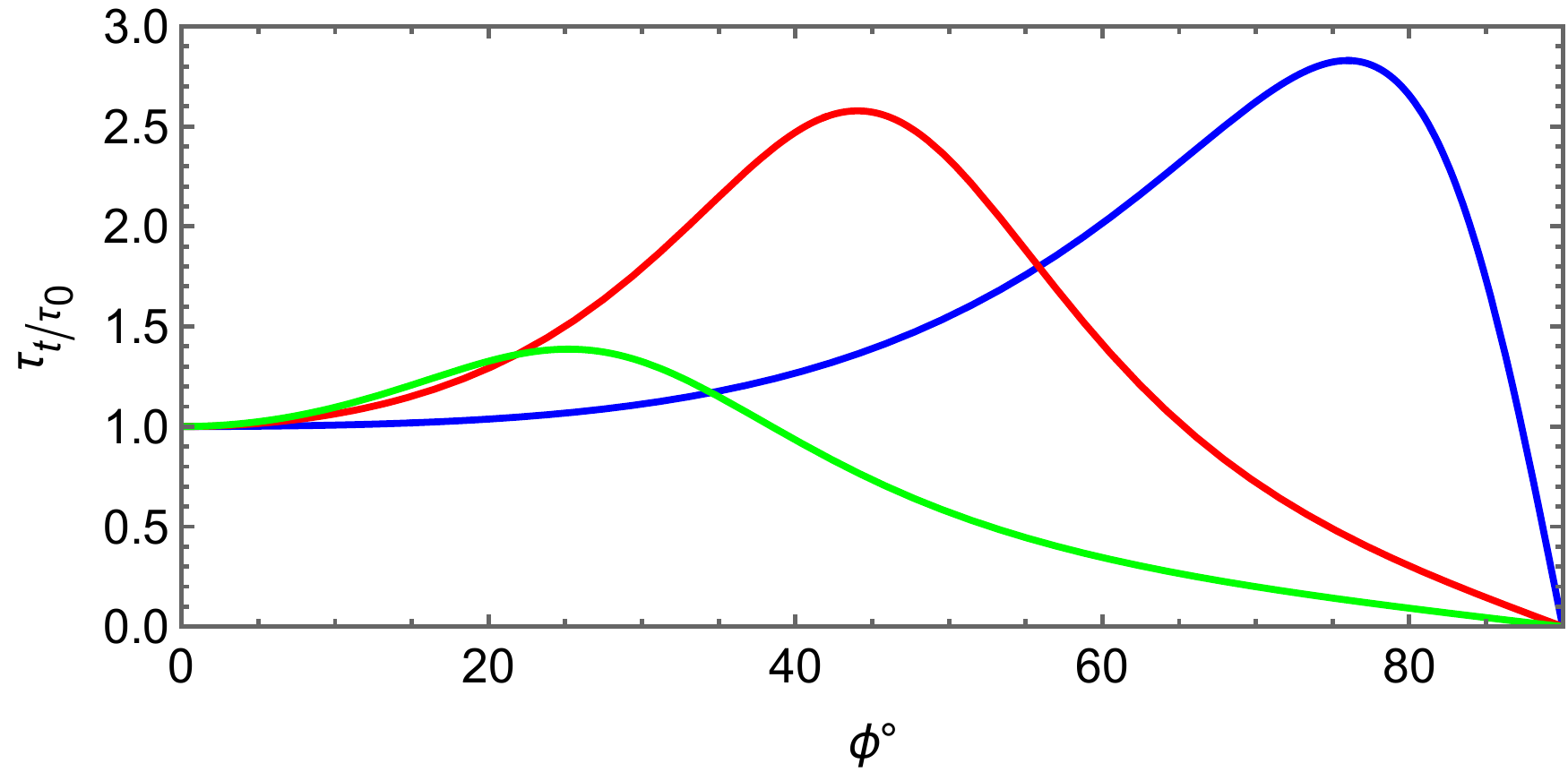}
  }  \\
\subfloat[]{
	\centering\includegraphics[scale=0.48]{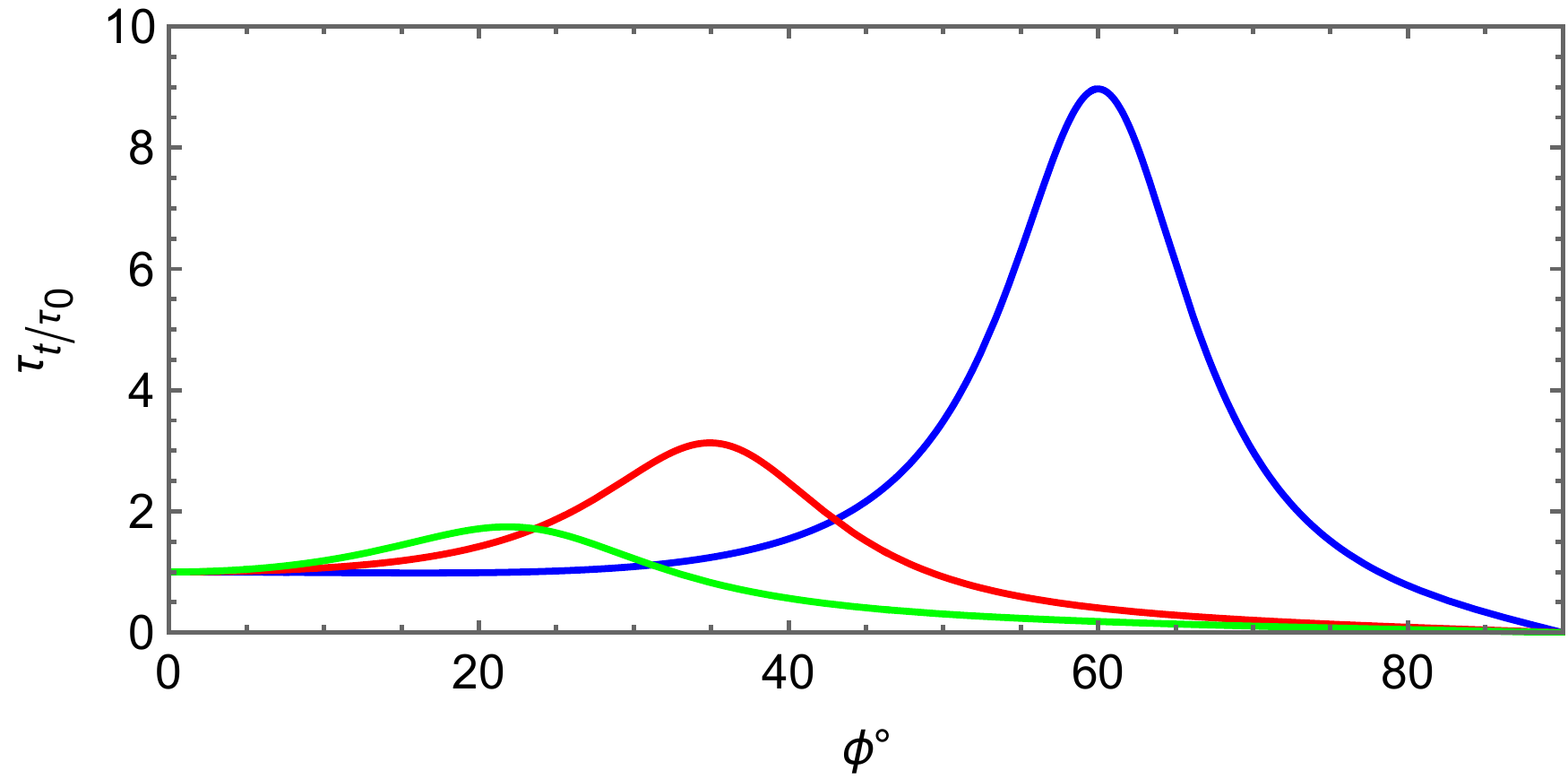}
} 
\subfloat[]{
	\centering\includegraphics[scale=0.48]{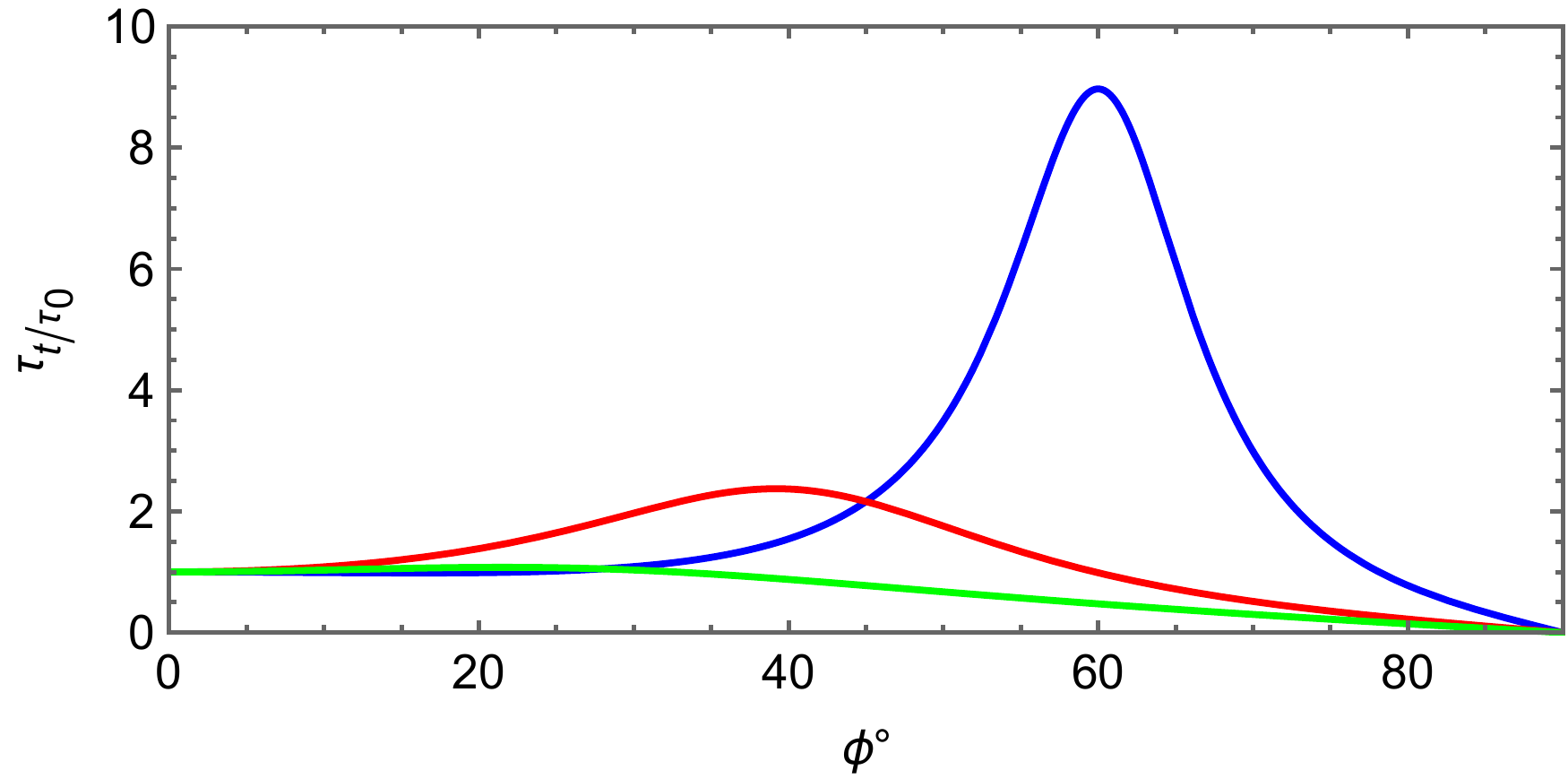}
} 
\caption{{(color online) The  group delay
time in transmission $\tau_{t}/\tau_{0}$ as a function of the
incident angle $\phi$ for $V_0=80$ meV, $d=80$~nm.
{\color{red}{(a)}}: $V_1=0$ meV and
{\color{red}{(b)}}: $V_1=20$ meV with
 $E=20$ meV (blue line)  $E=25$ meV (red line), $E=30$ meV (green
line).  
{\color{red}{(c)}}: $V_1=50$ meV, $E=30$ meV (blue line), $E=35$ meV (red line), $E=40$ meV (green line). {\color{red}{(d)}}:
$V_1=50$ meV, $E=30$ meV, $d=40$ nm (green line), $d=60$ nm (red line),  $d=80$ nm (blue line).}}
    \label{fig2}
\end{figure}

In Fig. \ref{fig2}, the group delay in transmission $\tau_{t}/\tau_{0}$
is shown versus the incident angle $\phi$ by choosing different values of the remaining 
physical parameters. 
It is evident that at normal incidence, i.e. $\phi=0$, the particles propagate through the barrier with the Fermi velocity $v_F$ ( $\tau_{t}/\tau_{0} = 1$), but that as $\phi$ increases, $\tau_{t}/\tau_{0} $  begins to progressively increase until it reaches a maximum, after which it decays exponentially. 
It approaches zero when $\phi=90^\circ$, since the wave vector inside the barrier becomes imaginary, and the wave function in the barrier region  becomes an evanescent wave. 
The behavior of $\tau_{t}/\tau_{0} $ is affected by incident energy $E$, barrier width $d$, and height $V_1$, as it drops as $E$ and $d$ increase, but increases as $V_1$ grows. This means that the linear potential can affect the group delay time by modulating it. 

\begin{figure}[H]
    \centering
    \subfloat[]{
        \centering
        \includegraphics[scale=0.48]{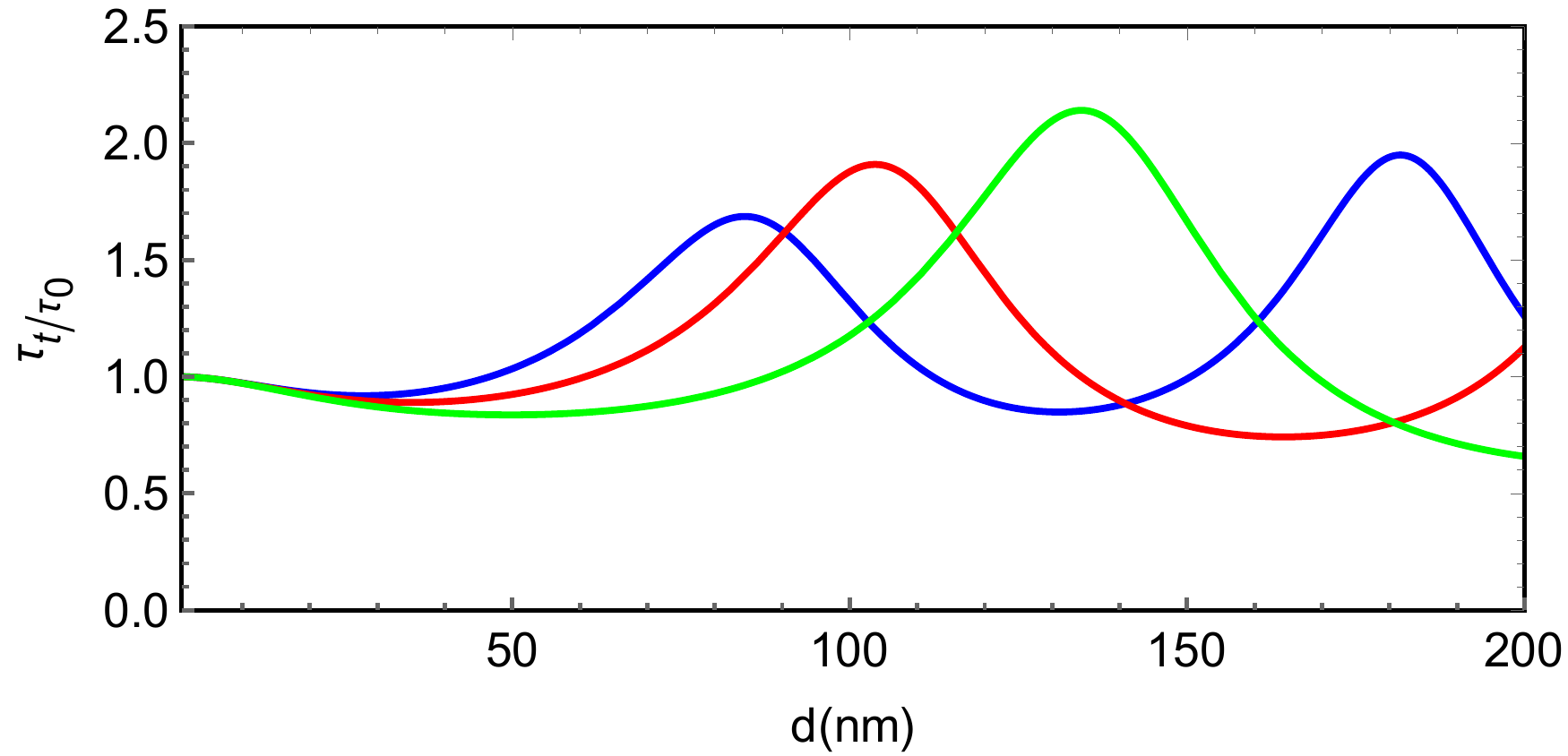}
        \label{fig7a}
    }
\subfloat[]{
        \centering
        \includegraphics[scale=0.48]{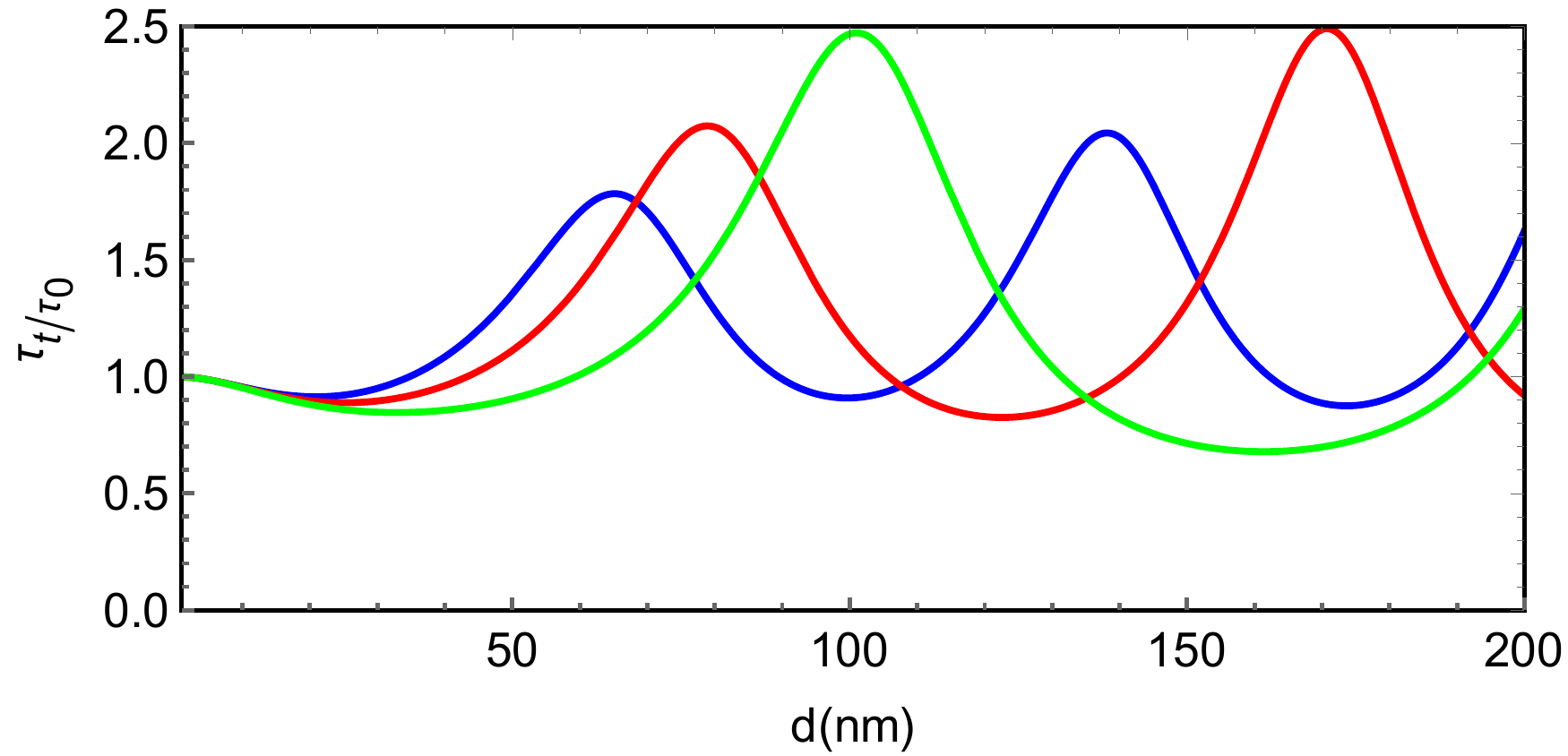}
        \label{fig7b}}\\
    \subfloat[]{
    \centering
    \includegraphics[scale=0.48]{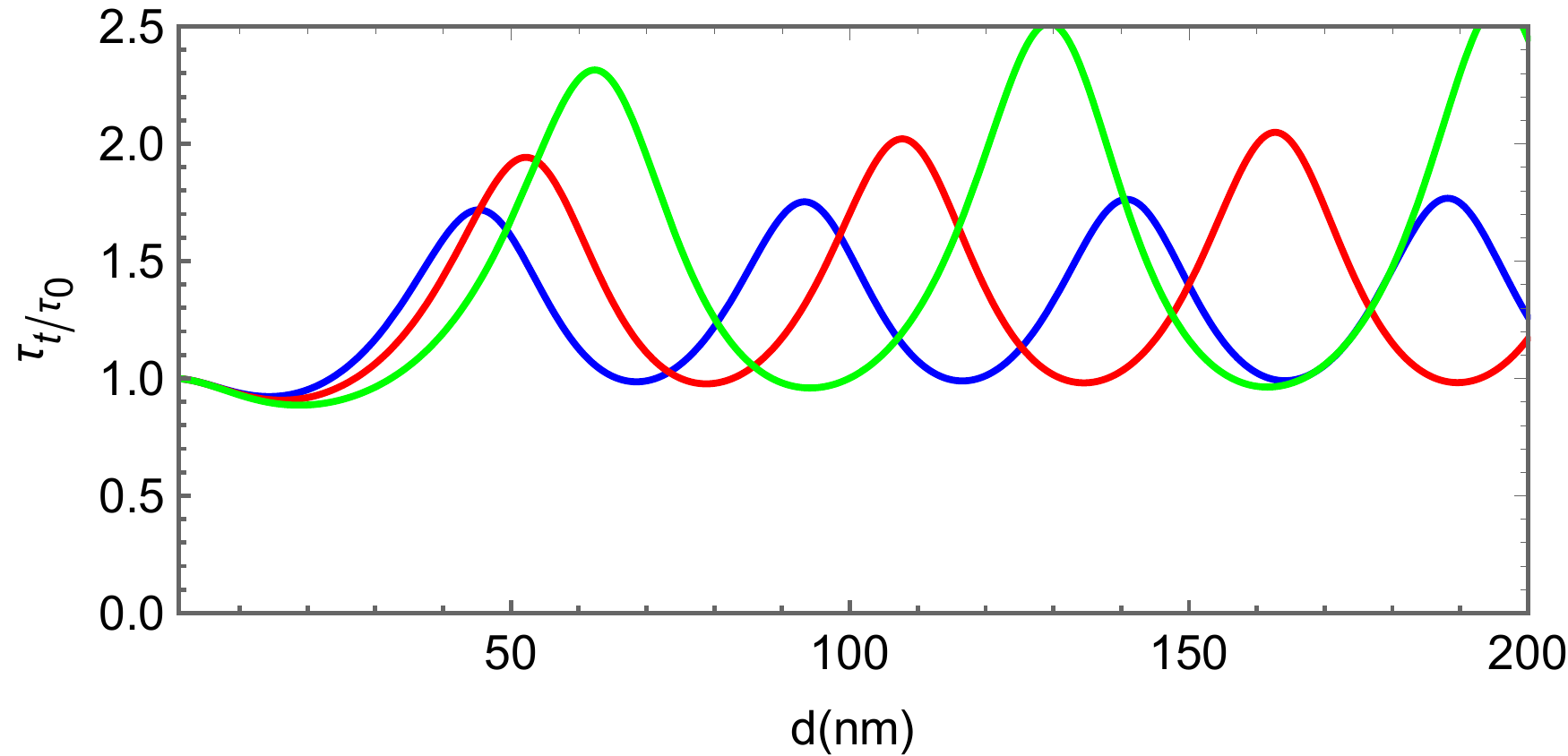}
    \label{fig7c}}
    \subfloat[]{
        \centering
        \includegraphics[scale=0.48]{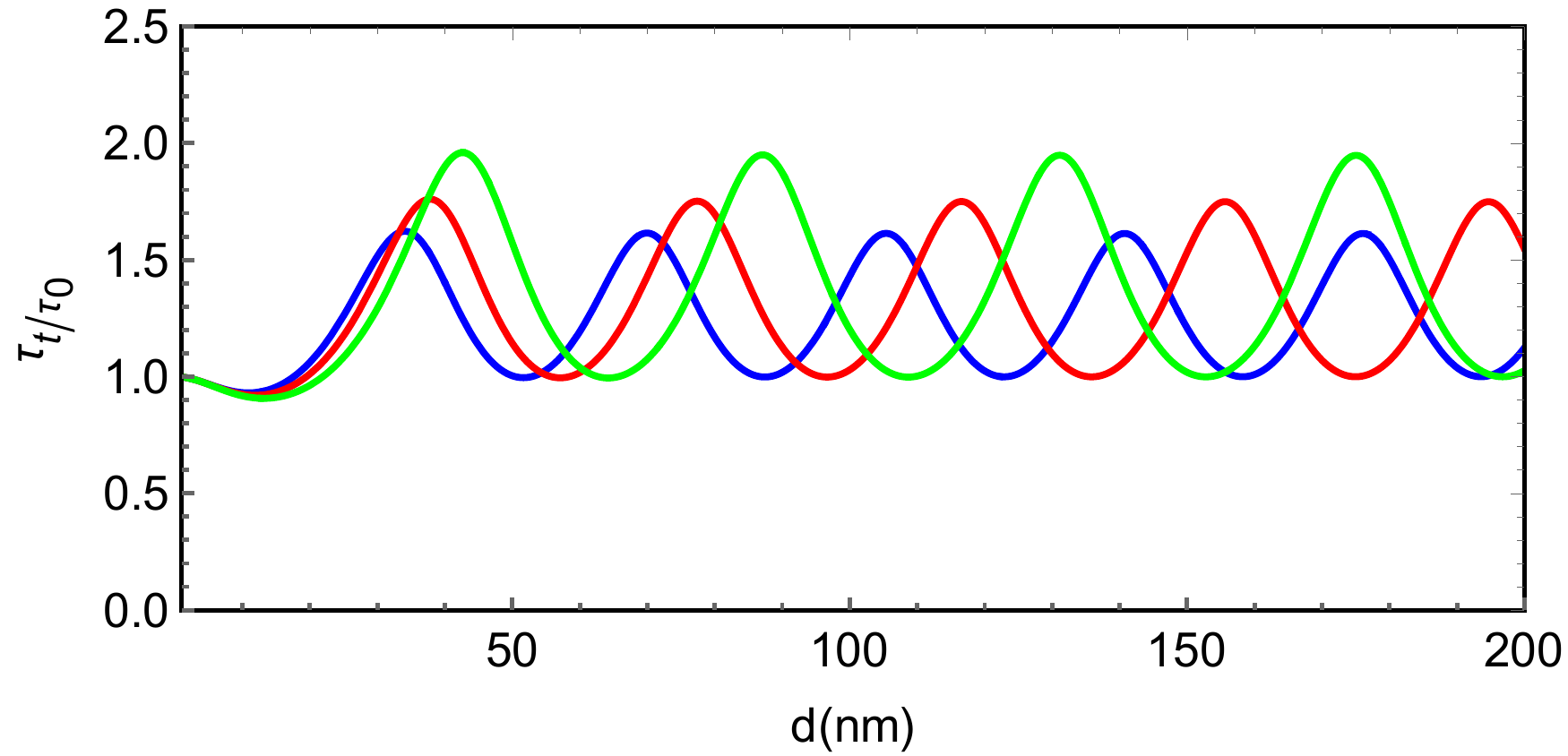}
        \label{fig7d}}
\caption{{(color online)  The group delay time in transmission $\tau_{t}/\tau_{0}$ as a function of the barrier width $d$ for $V_0=80$ meV, $\phi=30^{\circ}$, $E=20$ meV (blue line), $E=25$ meV (red line), $E=30$ meV (green
line). {\color{red}{(a)}}: $V_1=0$ meV, 
{\color{red}{(b)}}: $V_1=20$ meV, {\color{red}{(c)}}: $V_1=50$ meV and {\color{red}{(d)}}:
$V_1= V_0=80$ meV.}}
    \label{fig7}
\end{figure}

In Fig. \ref{fig7}, we plot the group delay time
$\tau_{t}/\tau_{0}$ as a function of  the barrier widths $d$ for
   the barrier heights  {\color{red}{(a)}}: $V_1=0$ meV,  {\color{red}{(b)}}:
$V_1=20$ meV,  {\color{red}{(c)}}: $V_1=50$ meV and 
 {\color{red}{(d)}}: $V_1= V_0$. The other computation parameters being $V_0=80$ meV,
$\phi=30^{\circ}$ and for different values of $E=20$ meV, $E=25$ meV  and $E=30$ meV. The possibility of modulating $\tau_{t}/\tau_{0}$ by changing the height of the potential barrier through different applied gate voltages is also present in the barrier tilting structure. Particularly interesting is that if  $V_0$ is kept constant and  $V_1$ is modified, $\tau_{t}/\tau_{0}$  shows identical behavior in Fig.
\ref{fig7a} for $V_1=0$ meV and Fig.
\ref{fig7b} for $V_1=20$ meV.
But if  $V_1$ is increased from $V_1=50$ meV up to $V_1= V_0$, the oscillations in  $\tau_{t}/\tau_{0}$  are again observed as
illustrated in Fig. \ref{fig7c} and Fig. \ref{fig7d} such that  their number 
increases 
and also their peak value decreases. 
On the other hand, we 
observe   $\tau_{t}/\tau_{0}$
increasing with the increase of incident energy $E$. Therefore the
incident energy modifies the period and amplitude of the
oscillations by increasing them. 
As the barrier width $ d $ is increased, the peaks show a discernible spread due to the Fabry-P\'erot enhancement. 
The particles travel back inside the tilting barrier due to Fabry-P\'erot resonances between the barrier edges, which explains
$\tau_{t}/\tau_{0}$.
 
\begin{figure}[H]
	\centering
	\subfloat[ ]{
		\centering
		\includegraphics[scale=0.48]{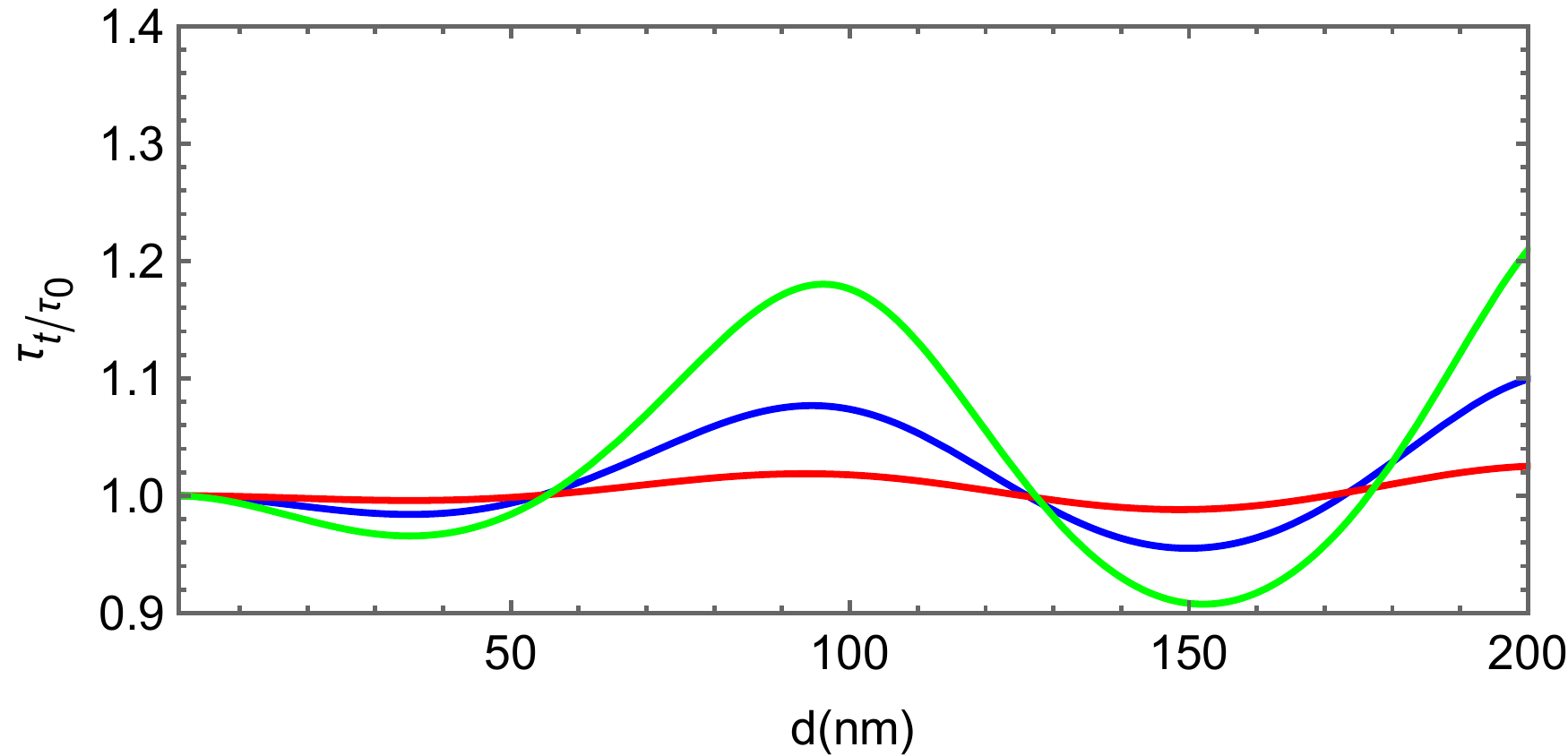}
		\label{fig8a}
	}
	\subfloat[]{
		\centering
		\includegraphics[scale=0.48]{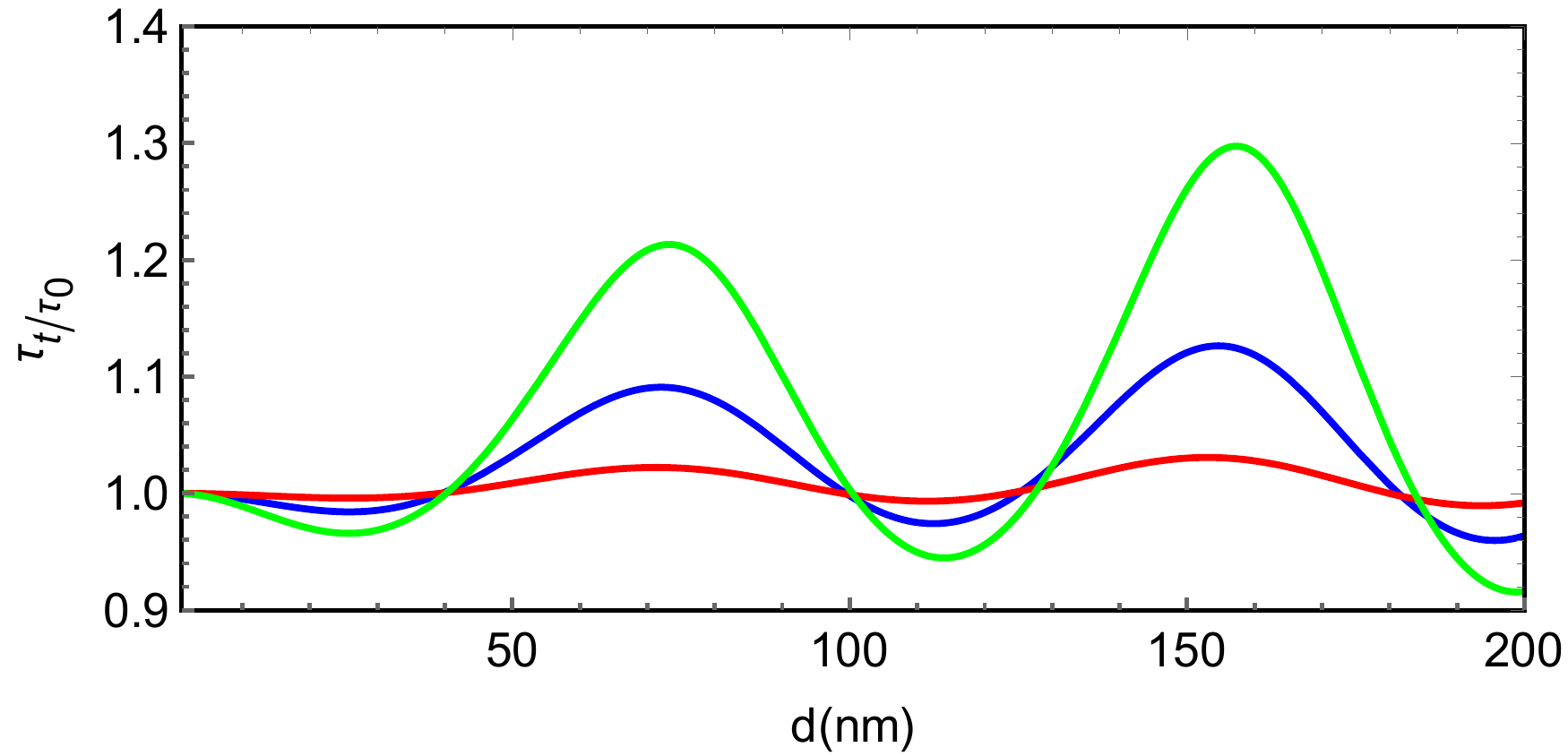}
		\label{fig8b}}\\
	\subfloat[]{
		\centering
		\includegraphics[scale=0.48]{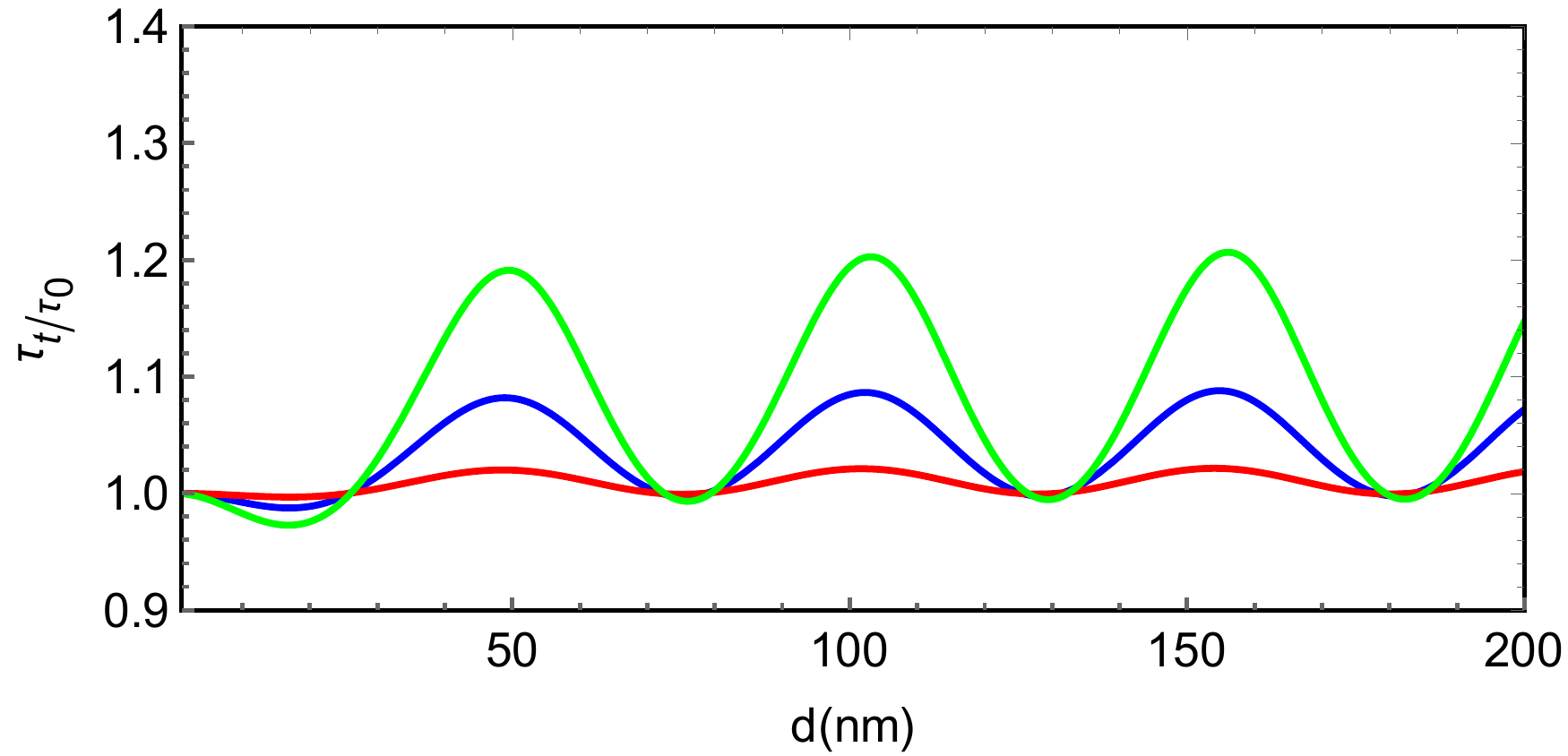}
		\label{fig8c}}
	\subfloat[]{
		\centering
		\includegraphics[scale=0.48]{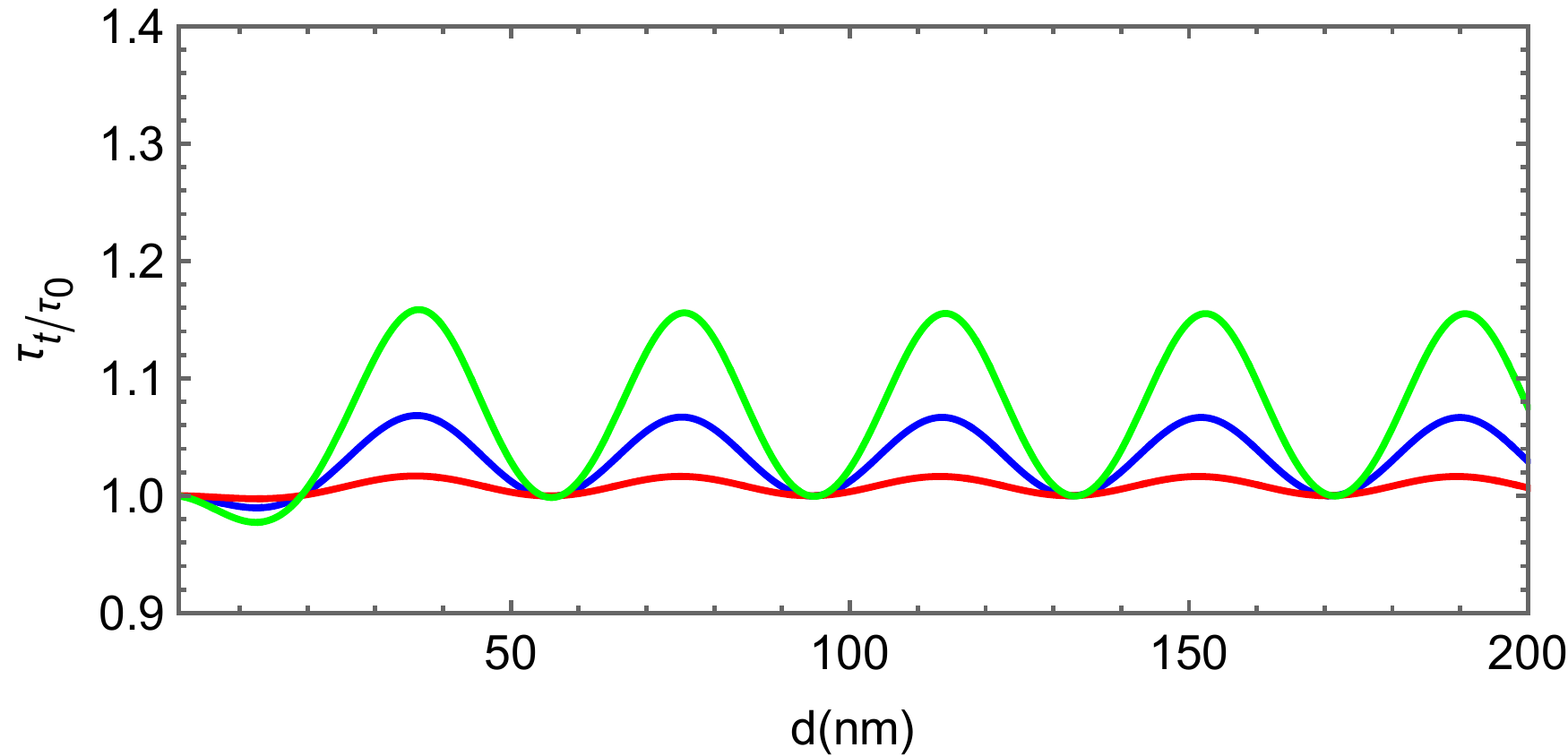}
		\label{fig8d}}
	\caption{{(color online) The group delay time
			in transmission $\tau_{t}/\tau_{0}$ as a function of the barrier
			width $d$ for $V_0=80$ meV, $E=30$ meV, $\phi=5^{\circ}$ (red
			line), $\phi=10^{\circ}$, (blue line),$\phi=15^{\circ}$ (green
			line). {\color{red}{(a)}}: $V_1=0$ meV,
			{\color{red}{(b)}}: $V_1=20$ meV, {\color{red}{(c)}}: $V_1=50$ meV, {\color{red}{(d)}}:
			$V_1\approx V_0=80$ meV.  }}
	\label{fig8}
\end{figure}

Fig. \ref{fig8} shows  the group delay time $\tau_{t}/\tau_{0}$ as a function of
the barrier width $d$ for different values of the incident angle
$\phi=5^{\circ}$ (red line), $\phi=10^{\circ}$ (blue line),
$\phi=15^{\circ}$ (green line), $E=30$ meV and we choose the remaining parameters as in Fig. \ref{fig7}. With the Fermi velocity $v_F$, the particles pass through the barrier, but when  $d$ increases the group delay time in
transmission $\tau_{t}/\tau_{0}$ starts to oscillates with peak
increasing. {Notice that for $V_1=50$ meV and  $V_1\approx V_0=80$ meV
	as presented, respectively,  Fig. \ref{fig8c} and  Fig. \ref{fig8d}, $\tau_{t}/\tau_{0}$ oscillates twice and  the peak value decreases compared to
Fig. \ref{fig8a} and Fig. \ref{fig8b}}, where $\tau_0$ is the time it would take a particle to travel the distance $d$ if the barrier did not exist.  Here we observe that 
the particles propagate through the barrier with the Fermi
velocity $v_F$ ($\tau_{t}/\tau_{0}= 1$).  When
$d$
increases one sees that   $\tau_{t}/\tau_{0}$ begins  to
oscillate  and the associated amplitude increases with the increase of the
incident angle $\phi$, while the peaks did not get influenced and they are still in the same positions.
When $v_F$ is equivalent to the speed of light $c$ in optics, the group delay time $\tau_{t}$ may be smaller than $\tau_{0}$ ($\tau_{t}/\tau_{0}<1$), indicating superluminality. This phenomena faster than light is relevant for the Hartman effect in the tunneling process.
On the other hand, we discover that particles propagate past the barrier at speeds greater than the Fermi velocity $v_F$.

\begin{figure}[H]
	\centering
	\subfloat[]{
		\centering
		\includegraphics[scale=0.34]{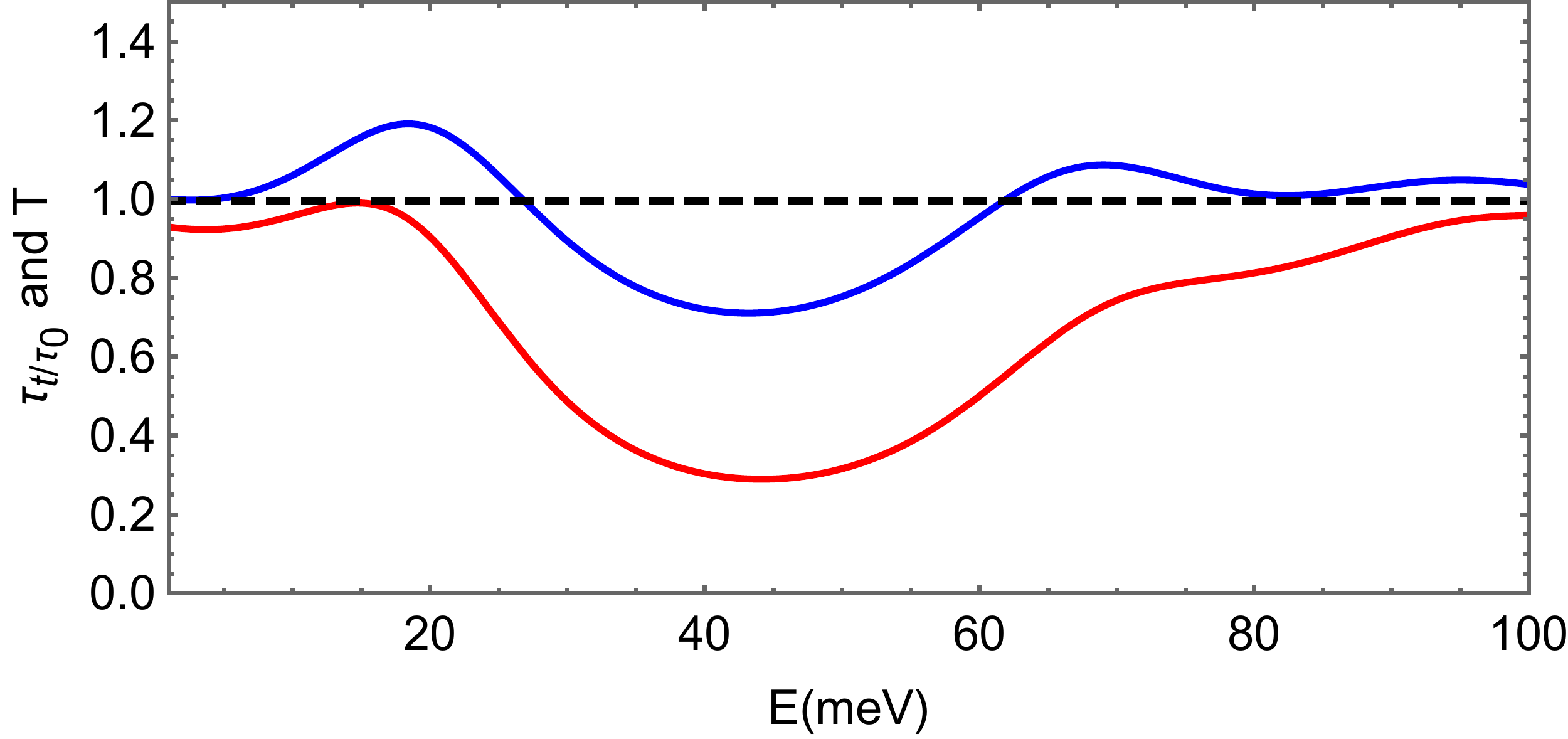}
		\label{fig5a}}
	\subfloat[]{
		\centering
		\includegraphics[scale=0.46]{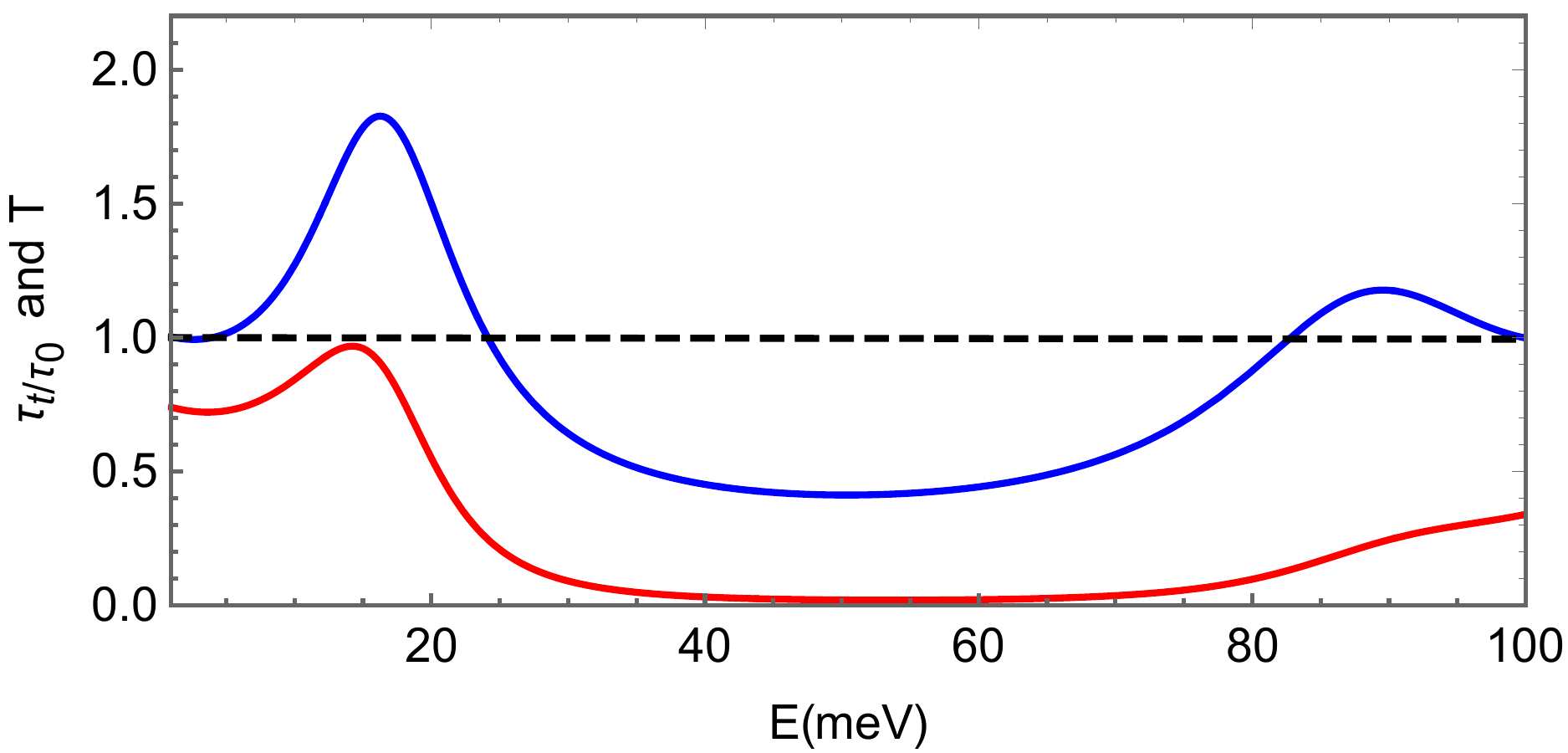}
		\label{fig5b}}
	\caption{{(color online) The
			group delay time 
			$\tau_{t}/\tau_{0}$ (blue line) and transmission probability $T$
			(red line) as a function of the incident energy $E$ at
			two incident angles {\color{red}{(a)}}:  $\phi=15^{\circ}$ and {\color{red}{(b)}}:  $\phi=30^{\circ}$, with
			$V_0=60$
			meV,  $V_1=20$ meV, $d=80$ nm.}}
	\label{fig5}
\end{figure}

In Fig. \ref{fig5} we plot the group delay time in transmission
$\tau_{t}/\tau_{0}$ (blue line) and transmission probability $T$
(red line) as a function of the incident energy $E$ for two values of  the incident angle  $\phi=15^{\circ}, 30^{\circ}$ in Fig. \ref{fig5a} and Fig. \ref{fig5b} 
 with $V_0=60$ meV,  $V_1=20$ meV and $d=80$ nm. 
 It can be seen that when $ E $ increases, both quantities exhibit closed behavior and they can be modulated by modifying 
 the incident angle. Additionally, we see oscillating behavior in the group delay time and transmission probability.
 This is due to the overlapping of the reflected and incident waves, which causes self-interference delay. 
Furthermore, we see a peak in the $\tau_{t}/\tau_{0}$ behavior,
 which increases in tandem with  the increase 
 in the incident angle. 
As shown in Fig. \ref{fig5a}, when $E< V_0-V_1$, $ T $ decays exponentially due to the wave function in the region of the barrier becoming an evanescent wave, and goes down to a minimum, then starts increasing again. However, $ T $ approaches zero before starting to increase again when $E> V_0-V_1$ as depicted in 
Fig.~\ref{fig5b}.

\begin{figure}[H]
	\centering
	\subfloat[]{
		\centering
		\includegraphics[scale=0.66]{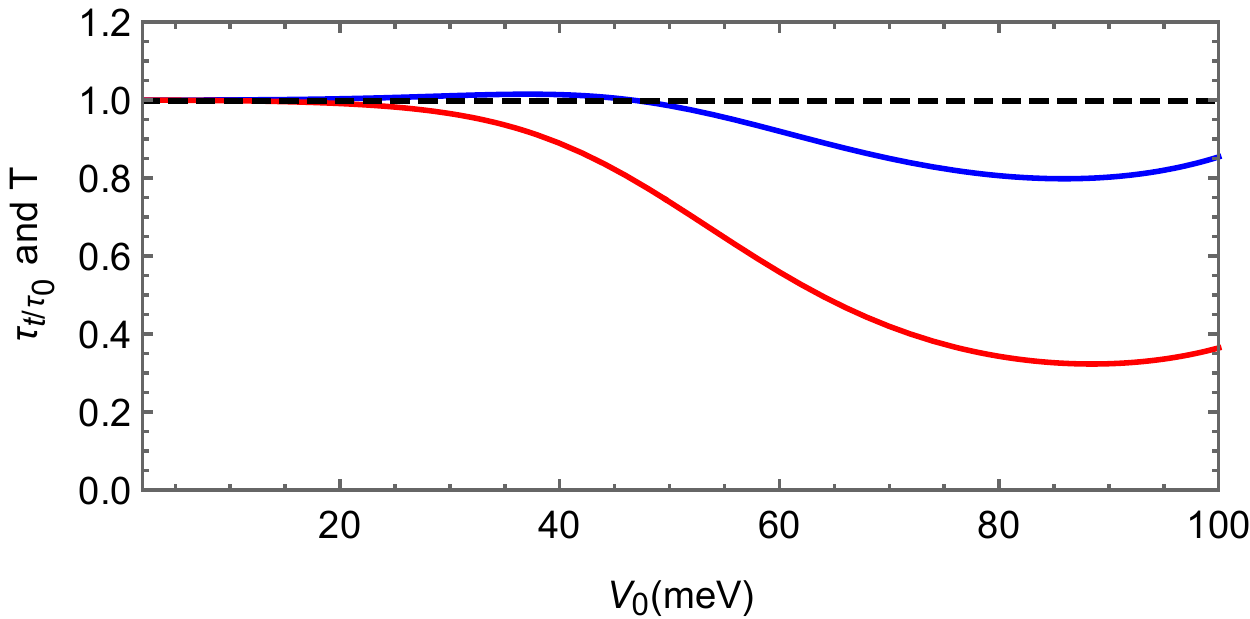}
		\label{fig6a}}
	\subfloat[]{
		\centering
		\includegraphics[scale=0.46]{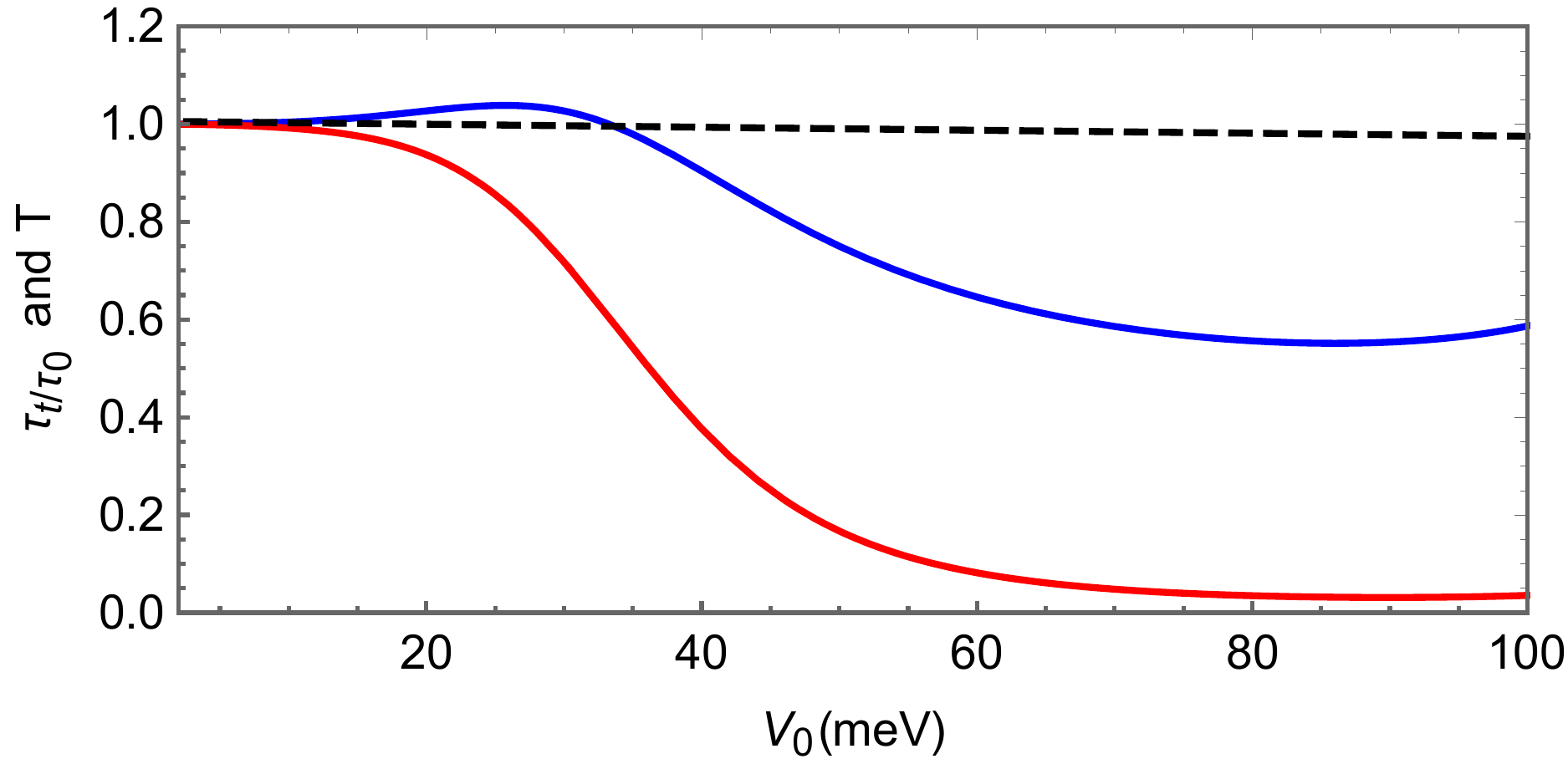}
		\label{fig6b}}
	\caption{{(color online) The
			group delay time 
			$\tau_{t}/\tau_{0}$ (blue line) and transmission probability $T$
			(red line) as a function of the barrier height $V_0$ at incident angles {\color{red}{(a)}}:  $\phi=15^{\circ}$ and {\color{red}{(b)}}:  $\phi=30^{\circ}$, with
			$E=50$
			meV, $V_1=0$ meV,  and $d=80$ nm.
	}}
	\label{fig6}
\end{figure}

 Fig. \ref{fig6} presents the group delay time  $\tau_{t}/\tau_{0}$ and the
transmission probability $T$ as a function of the barrier height $V_0$ for 
 $d=80$ nm, $E=50$ meV, and $V_1=0$ meV, with  Fig. \ref{fig6a} and Fig. \ref{fig6b} illustrate the case of   $\phi=15^{\circ}$ and $\phi=30^{\circ}$, respectively. 
For small values of $V_0$, one sees that the particles propagate through the tilted barrier at the Fermi velocity $v_F$, which they can transmit perfectly, $ T = 1 $. When $V_0$ is increased, $\tau_{t}/\tau_{0}$ and $T$ begin to rapidly fall toward a constant value that is independent of $V_0$.
We observe that the incident angle has an effect on $\tau_{t}/\tau_{0}$ and   $T$'s  behaviors since they decrease as it increases.
For $\phi=30^{\circ}$,  $ T $ stabilizes at zero regardless of the value of $ V_0 \geq 80$ meV, resulting in a total reflection. 

\begin{figure}[H]
	\centering
	\subfloat[]{
		\centering
		\includegraphics[scale=0.45]{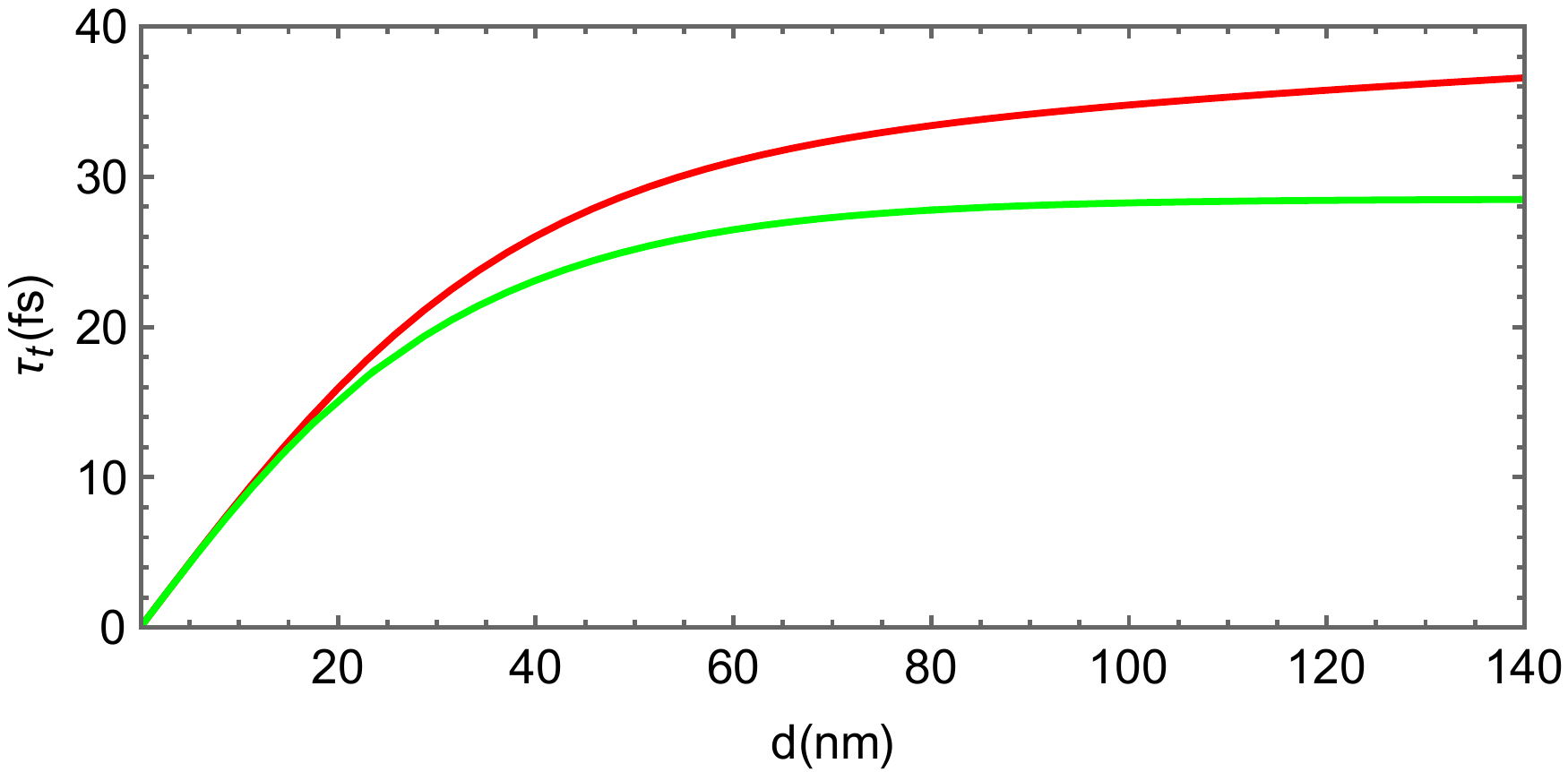}
		\label{fig9a}
	} \subfloat[]{
		\centering
		\includegraphics[scale=0.45]{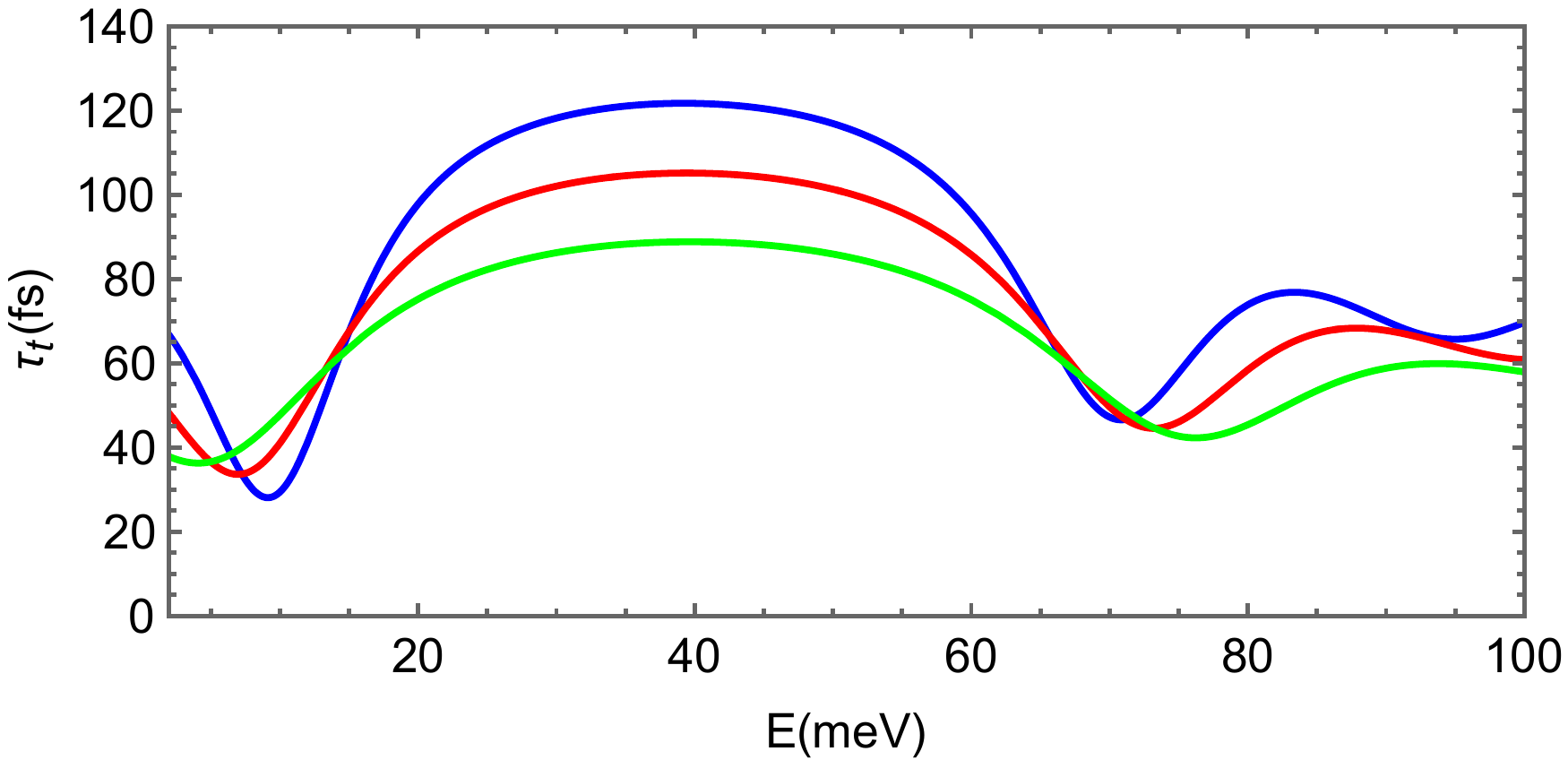}
		\label{fig9b}}
	\caption{{(color online) {\color{red}{(a)}}: The
			group delay time in transmission $\tau_{t}$ as a
			function of the barrier width $d$ for $V_0=40$ meV, $E=40$ meV,
			$\phi=30^{\circ}$, with two values $V_1=20$ meV (red line) and $V_1\approx V_0=40$
			meV (green line). {\color{red}{(b)}}: $\tau_{t}$ as a function of the incident
			energy $E$ for $V_0=40$ meV, $V_1=20$ meV, $\phi=30$, with three values  $d=90$ nm
			(blue line), $d=80$ nm (red line), $d=70$nm (green line). }}
	\label{fig9}
\end{figure}

Fig. \ref{fig9a} depicts the group delay time in
transmission $\tau_{t}$ as a function of the barrier width $d$ for
tilting barrier ($V_0=40$ meV, $V_1=20$ meV) as well as barrier square
$V_1= V_0=40$ meV with $E=40$ meV,
$\phi=30^{\circ}$.  
One notices that 
$\tau_{t}$ 
rapidly grows as $ d $ increases, eventually stabilizing at a maximum.
We get the same behavior shape with $ V_1=V_0 $, but with a decline. 
As a result, Fig.  \ref{fig9a} shows that the Hartmann effect exists at $V_1=
V_0=40$ meV, because $\tau_{t}$   saturates at a constant when the barrier width $ d $ is increased \cite{YueBan}. 
 Fig. \ref{fig9b} illustrates $\tau_{t}$ as a function of incident energy $ E $ for for $V_0=40$ meV, $V_1=20$ meV,
 $\phi=30^{\circ}$, demonstrating the impact of barrier widths
$d=90$ nm (blue line), $d=80$ nm (red
line) and $d=70$ nm (green line). 
We can see that $\tau_{t}$ oscillates as $ E $ increases, and that its peak increases as $ d $ increases.
When $ E $ grows, $\tau_{t}$ tends to settle at a certain value.

\section{Conclusion}

We investigated the group delay time in transmission $\tau_{t}/\tau_{0}$ for Dirac fermions in graphene scattered along the $ x $-axis by a linear barrier potential. 
The group delay time has been demonstrated to oscillate in response to several physical parameters such as barrier width $d$, incident angle $\phi$, incident energy $E$, and two barrier heights ($V_0,V_1$).   
When the barrier width $ d $ becomes large enough, our theoretical investigation supports the existence of group delay time saturation.
It also proves that quantum interference has a significant impact on particle tunneling in graphene via a tilted barrier.

Also we demonstrated that the physical parameters  that characterize our system can be used to modify the behavior of $\tau_t/\tau_0$.
Furthermore, we  discovered that the group delay time in transmission equals unity at certain critical values of incidence energy, incident angle, and barrier width, i.e. $\tau_{t}/\tau_{0}= 1$, implying that particles travel across the barrier with the Fermi velocity $v_F$.
Finally, we expect that all of the findings will be valuable not only for the theoretical research of the tunneling effect but also for graphene's technological applications.

\section*{Acknowledgment}

\end{document}